\documentclass[sigconf]{acmart}

\usepackage{booktabs} 

\copyrightyear{2017} 
\acmYear{2017} 
\setcopyright{acmcopyright}

\acmConference{CIKM'17}{}{November 6--10, 2017, Singapore.}
\acmPrice{15.00}
\acmDOI{http://dx.doi.org/10.1145/3132847.3132914}
\acmISBN{ISBN 978-1-4503-4918-5/17/11} 

\begin{document}

\title[DeepRank]{DeepRank: A New Deep Architecture for Relevance Ranking in Information Retrieval}
\author{Liang Pang${}^{\dag\ast}$, Yanyan Lan${}^{\dag\ast}$, Jiafeng Guo${}^{\dag\ast}$, Jun Xu${}^{\dag\ast}$, Jingfang Xu${}^{\ddag}$, Xueqi Cheng${}^{\dag\ast}$}
\email{pl8787@gmail.com, {lanyanyan, guojiafeng, junxu, cxq}@ict.ac.cn, xujingfang@sogou-inc.com}
\affiliation{
  \institution{
  $\dag$CAS Key Lab of Network Data Science and Technology, Institute of Computing Technology, \\ Chinese Academy of Sciences, Beijing, China\\
$\ast$University of Chinese Academy of Sciences, Beijing, China\\
  $\ddag$Sogou Inc, Beijing, China}
}
\renewcommand{\shortauthors}{L. Pang et al.}
\begin{abstract}
This paper concerns a deep learning approach to relevance ranking in information retrieval (IR). Existing deep IR models such as DSSM and CDSSM directly apply neural networks to generate ranking scores, without explicit understandings of the relevance. According to the human judgement process, a relevance label is generated by the following three steps: 1) relevant locations are detected; 2) local relevances are determined; 3) local relevances are aggregated to output the relevance label. In this paper we propose a new deep learning architecture, namely DeepRank, to simulate the above human judgment process. Firstly, a detection strategy is designed to extract the relevant contexts. Then, a measure network is applied to determine the local relevances by utilizing a convolutional neural network (CNN) or two-dimensional gated recurrent units (2D-GRU). Finally, an aggregation network with sequential integration and term gating mechanism is used to produce a global relevance score. DeepRank well captures important IR characteristics, including exact/semantic matching signals, proximity heuristics, query term importance, and diverse relevance requirement. Experiments on both benchmark LETOR dataset and a large scale clickthrough data show that DeepRank can significantly outperform learning to ranking methods, and existing deep learning methods.
\end{abstract}
\maketitle
\section{Introduction}\label{sec:introduction}
Relevance ranking is a core problem of information retrieval. Given a query and a set of candidate documents, a scoring function is usually utilized to determine the relevance degree of a document with respect to the query. Then a ranking list is produced by sorting in descending order of the relevance score. Modern learning to rank approach applies machine learning techniques to the ranking function, which combines different kinds of human knowledge (i.e.~relevance features such as BM25~\cite{robertson1994some} and PageRank~\cite{page1999pagerank}) and therefore has achieved great improvements on the ranking performances~\cite{liu2009learning}. However, a successful learning to rank algorithm usually relies on effective handcrafted features for the learning process. The feature engineering work is usually time-consuming, incomplete and over-specified, which largely hinder the further development of this approach~\cite{DRMM}.

Recently, deep learning approach~\cite{lecun2015deep} has shown great success in many machine learning applications such as speech recognition, computer vision, and natural language processing (NLP), owing to their ability of automatically learning the effective data representations (features). Therefore, a new direction of Neural IR is proposed to resort to deep learning for tackling the feature engineering problem of learning to rank, by directly using only automatically learned features from raw text of query and document. There have been some pioneer work, including DSSM~\cite{DSSM}, CDSSM~\cite{CDSSM}, and DRMM~\cite{DRMM}. Both DSSM and CDSSM directly apply deep neural networks to obtain the semantic representations of query and document, and the ranking score is produced by computing their cosine similarity. Guo et al.~\cite{DRMM} argued that DSSM and CDSSM only consider the semantic matching between query and document, but ignore the more important relevance matching characteristics, such as exact matching signals, query term importance, and diverse matching requirement~\cite{robertson1994some}. Thus they proposed another deep architecture, i.e.~DRMM, to solve this problem. However, DRMM does not explicitly model the relevance generation process, and fails to capture important IR characteristics such as passage retrieval intrinsics~\cite{lv2009positional} and proximity heuristics~\cite{tao2007exploration}.  

Inspired by the human judgement process, we propose a new deep learning architecture, namely DeepRank, to better capture the relevance intrinsics. According to the illustration in \cite{wu2007retrospective}, the human judgement process can be divided into three steps. Human annotators first scan the whole document to detect the relevant locations. Then the local relevance for each detected location is decided. Finally, those local relevances are combined to form the global relevance of the entire document. Consequently, DeepRank contains three parts to simulate the human judgement process, by tackling the following three problems:

{\em Where does the relevance occur?}  According to the query-centric assumption proposed in~\cite{wu2007retrospective}, the relevant information for a query only locates in the contexts around query terms. Therefore, the context with a query term at the center position, namely query-centric context, is recognized as the relevant location in the detection step.

{\em How to measure the local relevance?} After the detection step, a measure network is utilized to determine the local relevance between query and each query-centric context. Firstly, a tensor is constructed to incorporate both the word representations of query/query-centric context, and the interactions between them. Then a CNN or 2D-GRU is applied on the tensor to output the representation of the local relevance. In this way, important IR characteristics such as exact/semantic matching signals, passage retrieval intrinsics, and proximity heuristics can be well captured.

{\em How to aggregate such local relevances to determine the global relevance score?} As shown by \cite{wu2007retrospective}, two factors are important for user\rq{}s complex principles of aggregating local relevances, i.e.~query term importance \cite{fang2004formal} and diverse relevance requirement \cite{robertson1994some}. Therefore we propose to first aggregate local relevances at query term level, and then make the combination by considering weights of different terms, via a term gating network. To obtain the term level relevance, we first group the query-centric contexts with the same central word together. Then a recurrent neural network (RNN) such as GRU~\cite{GRU_cho2014learning} or LSTM~\cite{LSTM_graves2013speech} is utilized to aggregate such local relevances sequentially, by further considering the position information of these query-centric contexts in the whole document.

Above all, DeepRank is a new architecture composed of three components, i.e.~a detection strategy, a measure network with CNN/2D-GRU, and an aggregation network with term gating and RNN inside. Therefore, DeepRank can be trained end-to-end with the pairwise ranking loss via stochastic gradient descent. We conduct experiments on both benchmark LETOR4.0 data and a large scale clickthrough data collected from a commercial search engine. The experimental results show that: 1) Existing deep IR methods such as DSSM, CDSSM, and DRMM perform much worse, if using only automatically learned features, than the pairwise and listwise learning to rank methods. 2) DeepRank significantly outperforms not only all the existing deep IR models but also all the pairwise and listwise learning to rank baseline methods. 3) If we incorporate handcrafted features into the model, as did in SQA \cite{SQA}, DeepRank will be further improved, and the performance is better than SQA. We also conduct a detailed experimental analysis on DeepRank to investigate the influences of different settings. 

To the best of our knowledge, DeepRank is the first deep IR model to outperform existing learning to rank models.

\section{Related Work}\label{sec:relatedworks}
We first review related work on relevance ranking for IR, including learning to rank methods and deep learning methods.

\subsection{Learning to Rank Methods}
In the past few decades, machine learning techniques have been applied to IR, and gained great improvements to this area. This direction is called learning to rank. Major learning to rank methods can be grouped into three categories: pointwise, pairwise and listwise approach. Different approaches define different input and output spaces, use different hypotheses, and employ different loss functions. Pointwise approach, such as logistic regression~\cite{gey1994inferring}, inputs a feature vector of each single document and outputs the relevance degree of each single document. Pairwise approach, such as RankSVM~\cite{RankSVM} and RankBoost~\cite{RankBoost}, inputs pairs of documents, both represented by feature vectors and outputs the pairwise preference between each pair of documents. Listwise approach, such as ListNet~\cite{ListNet}, AdaRank~\cite{AdaRank} and LambdaMart~\cite{burges2010ranknet}, inputs a set of document features associated with query and outputs the ranked list. All these approaches focus on learning the optimal way of combining features through discriminative training. However, a successful learning to rank algorithm relies on effective handcrafted features for the learning process. The feature engineering work is usually time-consuming, incomplete and over-specified, which largely hinder the further development of this direction~\cite{DRMM}.

\subsection{Deep Learning Methods}
Recently, deep learning techniques have been applied to IR, for automatically learning effective ranking features. Examples include DSSM~\cite{DSSM}, CDSSM~\cite{CDSSM}, and DRMM~\cite{DRMM}. 
DSSM uses a deep neural network (DNN) to map both query and document to a common semantic space. Then the relevance score is calculated as the cosine similarity between these two vectors. Rather than using DNN, CDSSM is proposed to use CNN to better preserve the local word order information when capturing contextual information of query/document. Then max-pooling strategies are adopted to filter the salient semantic concepts to form a sentence level representation. However, DSSM and CDSSM view IR as a semantic matching problem, and focus on generating a good sentence level representations for query and document. They ignore the important intrinsics of relevance ranking in IR. Guo et al.~\cite{DRMM} first point out the differences between semantic matching and relevance matching. They propose a new deep learning architecture DRMM to model IR\rq{}s own characteristics, including exact matching signals, query terms importance, and diverse matching requirement. Specifically, DRMM first builds local interactions between each pair of words from query and document based on word embeddings, and then maps the local interactions to a matching histogram for each query term. Then DRMM employs DNN to learn hierarchical matching patterns. Finally, the relevance score is generated by aggregating the term level scores via a term gating network. Though DRMM has made the first step to design deep learning model specially for IR, it does not explicitly model the relevance generation process of human. It also fails to model the important IR intrinsics such as passage retrieval strategies and proximity heuristics.

Another related sort of deep models, flourishing in NLP, provide a new way of thinking if we treat IR task as a general text matching task, i.e.~query matches document. These work can be mainly categorized as representation focused models and interaction focused models. The representation focused models try to build a good representation for each single text with a neural network, and then conduct matching between the two vectors. The DSSM and CDSSM mentioned above belong to this category. There are also some other ones such as ARC-I~\cite{ARC-II} model, which builds on word embeddings and makes use of convolutional layers and pooling layers to extract compositional text representation. The interaction focused models first build the local interactions between two texts, and then use neural networks to learn the more complicated interaction patterns for matching. Typical examples include ARC-II~\cite{ARC-II} and MatchPyramid~\cite{MatchPyramid, pang2016study}, and Match-SRNN~\cite{Match-SRNN}. These models have been shown effective in text matching tasks such as paraphrase identification and question answering. DRMM can also be viewed as an interaction focused model.

\section{Motivation}\label{sec:motivation}
The motivation of our deep architecture comes from the process of human relevance judgement. As illustrated by \cite{wu2007retrospective}, a human annotator will first examine the whole document for local relevance information. Scanning involves iterating through every document location, and deciding for each whether local relevance information is found. As suggested by \cite{wu2007retrospective}, if one can find relevant information in a document, the relevant information must locate around the query terms inside the document, namely query-centric assumption. Finally, the local relevance for each query-centric context is combined to form the relevance of the entire document. Figure~\ref{judge} gives an example to describe the process of human relevance judgement, where user\rq{}s information need is represented by the query `Hubble Telescope Achievements\rq{}. For a given document, annotators will directly extract several contexts containing the keyword `Telescope\rq{} and `Hubble\rq{}, and determine whether they are relevant to the query. Then these local relevance information will be considered together to determine the global relevance label of the document with respect to the query.

Therefore, the relevance label is generated following three steps in the human judgement process: 1) detection of relevant locations; 2) measurement of local relevance; 3) aggregation of local relevances. Consequently, three problems are going to be tackled if we want to well capture the relevance: (1) Where does the relevance occur? (2) How to measure the local relevance? (3) How to aggregate such local relevance to determine the final relevance label? Accordingly, we design our deep learning architecture, namely DeepRank, to model the above relevance process.
\section{DeepRank}\label{sec:model}
\begin{figure*}[hbt]
  \centering
  \includegraphics[width=0.85\textwidth]{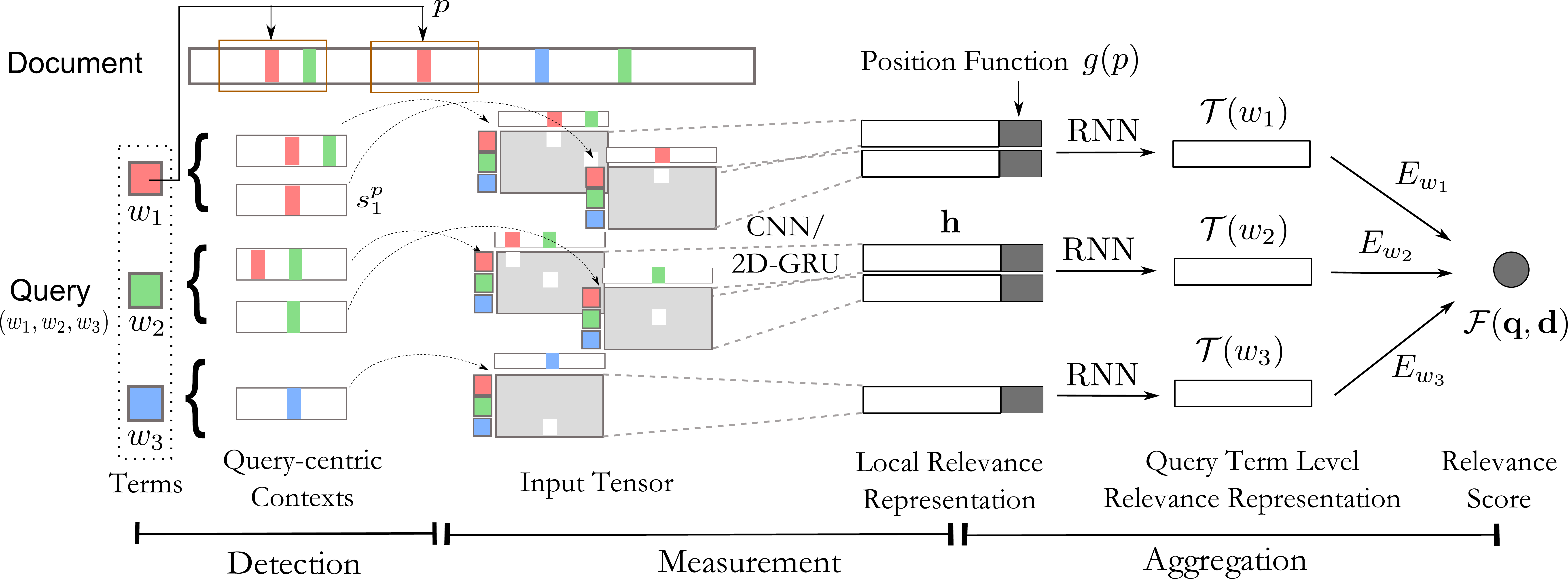}
  \caption{An illustration of DeepRank. }\label{model}
\end{figure*}
In this section, we demonstrate a new deep learning architecture for IR, namely DeepRank. DeepRank includes three parts to tackle the above three problems: {\em a Detection Strategy}, {\em a Measure Network}, and {\em an Aggregation Network}. In the detection step, the query-centric contexts are extracted to represent where the relevance occur. In the measurement step, CNN or 2D-GRU is adopted to measure the local relevance between query and each query-centric context. Finally in the aggregation step, RNN and a term gating network are utilized to aggregate those local relevances to a global one for ranking. Figure~\ref{model} gives an illustration of DeepRank.

We will first give some mathematical notations. Each query and document are represented as a sequence of word $\mathbf{q}=(w_1, \dots, w_M)$ and $\mathbf{d}=(v_1, \dots, v_N)$, where $w_i$ denotes the $i$-th word in the query and $v_j$ denotes the $j$-th word in the document, respectively. Consequently, we can use $\mathbf{d}_{(k)}[p]=(v_{p-k}, \cdots, v_p,\cdots,v_{p+k})$ to denote a piece of continuous text within a document, centered on the $p$-th word with the sequence length $2k+1$. If we use word2vec technique~\cite{mikolov2013distributed} to represent each word to a vector, each word sequence can be represented as a matrix. Taking query $\mathbf{q}=(w_1, \dots, w_M)$ for example, if the word embedding of $w_i$ is $\mathbf{x}_i$, the query will be represented as $\mathbf{Q}=[\mathbf{x_1}, \dots, \mathbf{x_M}]$, where each column stands for a word vector.

\subsection{Detection Strategy}
According to the query-centric assumption \cite{wu2007retrospective}, the relevance usually occurs at the locations where the query terms appear in the documents. Similar observations have also been obtained by the eye-tracking studies~\cite{eickhoff2015eye}, showing that annotators focus more on the location of query terms when they are scanning the whole document for relevance judgment. Therefore, we define the query-centric context as the relevant location, which is a context with a query term at the center position. Mathematically, given a query $\mathbf{q}$ and document $\mathbf{d}$, if query term $w_u$ appears at the $p$-th position in the document, i.e.~$w_u=v_p$, the query-centric context centered on this query term is represented as $\mathbf{s}_u^p(k)=\mathbf{d}_{k}[p]$.
After the detection step, we can obtain a set of word sequences $\{\mathbf{s}_u^p(k)\}$, with each one represents a local relevant location.
\begin{figure}[hbt]
  \centering
  \includegraphics[width=0.48\textwidth]{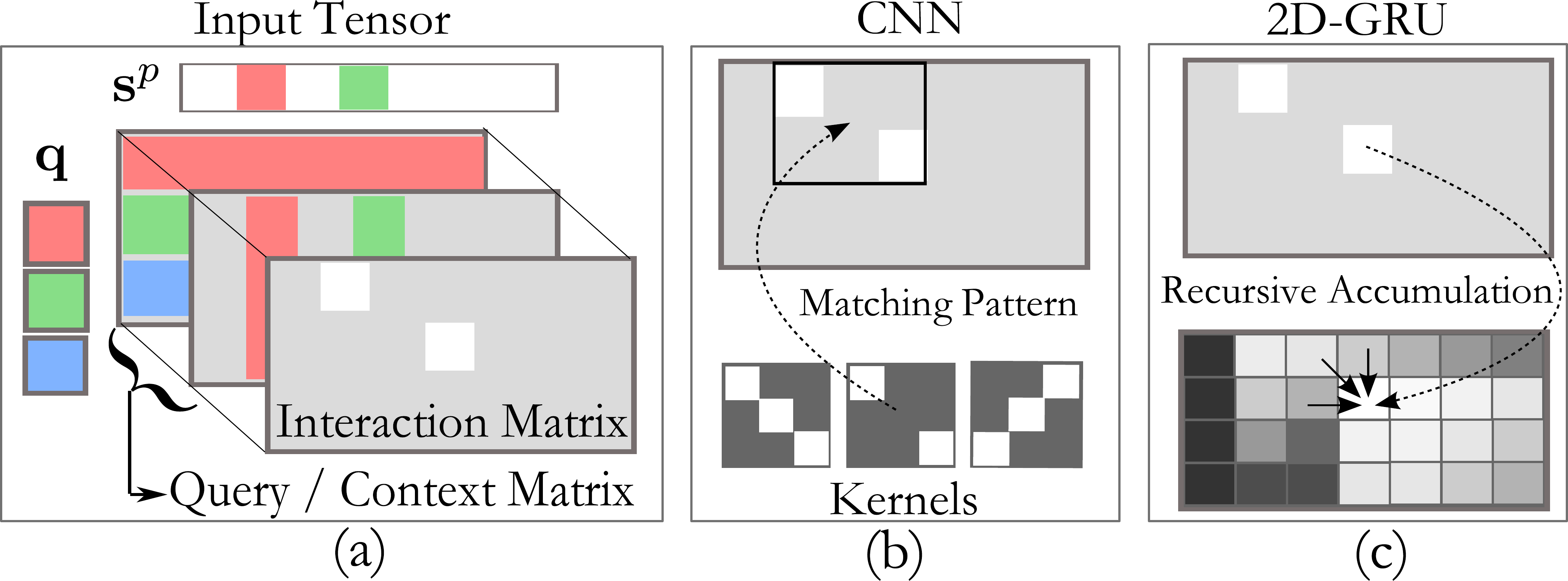}
  \caption{An illustration of measure network for a query and a query-centric context: (a) Input Tensor; (b) CNN; (c) 2D-GRU.}\label{fig.measure}
\end{figure}
\subsection{Measure Network}
The goal of the measurement step is to determine the local relevance, i.e.~relevance between query and each query-centric context.
As reviewed in Section~\ref{sec:relatedworks}, previous deep learning models for text matching can be divided into representation focused methods and interaction focused methods. The representation focused methods focus on extracting high level semantic representations of texts, starting from basic word representations. While the interaction focused methods turn to directly model the interactions between the two texts, starting from word-level interaction signals. In order to combine the best of both approaches, we propose a new measure network, as shown in Figure~\ref{fig.measure}. Firstly a tensor is constructed as the input. Then CNN or 2D-GRU is applied on the tensor to output a vector, which stands for the representation of local relevance. In this way, the important IR characteristics such as exact/semantic matching signals, passage retrieval intrinsics, and proximity heuristics can be well captured in the measurement step.
\subsubsection{Input Tensor}
The key idea of this layer is to feed both the word representations of query/query-centric context and the interactions between them into the input of the measure network. Specifically for a given query $\mathbf{q}$ and query-centric context $\mathbf{s}_u^p(k)$ with $w_u=v_p$, we denote the word-level interaction matrix used in MatchPyramid~\cite{MatchPyramid} and Match-SRNN~\cite{Match-SRNN} as $\bf{S}$, where each element $S_{ij}$ is defined as the similarity of corresponding words $w_i$ and $v_j$. For example, indicator function or cosine similarity can be used to capture the word-level exact or semantic matching signals, respectively. The mathematical formulas are shown as follows.
\begin{align}\label{Eq.matchmatix}
S_{ij}^{\text{ind}}=1 \,\,\text{if} \,\,w_i=v_j, \,\,\,S_{ij}^{\text{ind}}=0 \,\,\text{otherwise},\\
S_{ij}^{\text{cos}} = \mathbf{x_i}^T \mathbf{y_j} / (\|\mathbf{x_i}\| \cdot \|\mathbf{y_j}\|),
\end{align}
where $\mathbf{x_i}$ and $\mathbf{y_j}$ denote the word embeddings of $w_i$ and $v_j$, respectively.
To further incorporate the word representations of query/query-centric context to the input, we extend each element of $S_{ij}$ to a three-dimensional vector ${\bf\tilde S}_{ij}=[x_i,y_j,S_{ij}]^T$. Therefore, the original matrix $\bf{S}$ will become a three-order tensor, denoted as $\mathcal{S}$. In this way, the input tensor can be viewed as a combination of three matrices, i.e.~query matrix, query-centric context matrix, and word-level interaction matrix.

Based on the input tensor $\mathcal{S}$, various neural networks can be directly applied to obtain the representations for the local relevance between query and query-centric context. In this paper, we choose the CNN architecture in MatchPyramid and 2D-GRU architecture in Match-SRNN, mainly because they have the ability to capture important proximity heuristics for IR.
\subsubsection{Convolutional Neural Network} In this paper, we use a one-layer CNN in the measurement step, which includes a convolution operation and a max-pooling operation defined as follows. The convolution operation can extract various matching patterns from the input tensor $\mathcal{S}$, by using different kinds of kernels, as shown in Figure~\ref{fig.measure}. Then a max-pooling operation is used to filter significant matching patterns for further relevance determination. 
\begin{equation}\label{CNN}
	\begin{aligned}
		h_{i,j}^{(\kappa)}&= \sum_{l=1}^{3}\sum_{s=0}^{\gamma-1}\sum_{t=0}^{\gamma-1} w_{s,t}^{(\kappa)}{\cdot} \mathcal{S}_{i+s, j+t}^{(l)} {+} b^{(\kappa)},\\
		{h}^{(\kappa)} &= \max_{i,j}h_{i,j}^{(\kappa)},\,\,\kappa=1,\cdots,K\\
	\end{aligned}
\end{equation}
where $l$ denotes the $l$-th slide of the tensor, $\gamma$ denotes the fixed size of $K$ different kernels, $\mathcal{S}_{i+s,j+t}^{(l)}$ denote the $(i+s,j+t)$ element of the $l$-th matrix of the input tensor $\mathcal{S}$, $w_{s,t}^{\kappa}$ and $b^{\kappa}$ denotes parameters. 

Finally, all the significant matching patterns obtained from different kernels are concatenated to form a vector, i.e.~${\bf h}=[h^{(1)},\cdots,h^{(K)}]^T$, to represent the local relevance. This vector will be treated as the input to the aggregation network.
\subsubsection{Two-Dimensional Gated Recurrent Units}
Rather than using a hierarchical structure to capture the matching patterns, 2D-GRU in Match-SRNN~\cite{Match-SRNN} adopts a different sequential model to accumulate the matching signals. It is an extension of GRU~\cite{GRU_cho2014learning} (a typical variant of RNN) to two-dimensional data like matrix or tensor\footnote{Strictly speaking, tensor is not a two-dimensional representation. However, 2D-GRU can be directly applied on tensor by treating each element of the matrix as a vector.}. Specifically, 2D-GRU scans from top-left to bottom-right (or in a bidirectional way) recursively. At each position, the hidden representation depends on the representations of the top, left, diagonal and current positions in the matrix. The mathematical formulas are shown as follows.
\begin{equation}\label{2D-GRU}
	\begin{aligned}
		&\mathbf{c}=[\mathbf{h}_{i-1,j}^T,\mathbf{h}_{i,j-1}^T,\mathbf{h}_{i-1,j-1}^T,\mathcal{S}_{ij}^T]^T,\\
		&\mathbf{r}_\theta =\sigma(\mathbf{W}^{(r_\theta)}\mathbf{c}+\mathbf{b}^{(r_\theta)}), \,\, \theta = l,t,d,\\
		&\mathbf{z}'_{\phi} = \mathbf{W}^{(z_\phi)}\mathbf{c}+\mathbf{b}^{(z_\phi)}, \,\,\, \phi = m,l,t,d,\\
		&\mathbf{r}  =[\mathbf{r}_l^T,\mathbf{r}_t^T,\mathbf{r}_d^T]^T, [\mathbf{z}_m, \mathbf{z}_l, \mathbf{z}_t, \mathbf{z}_{d}] = \textrm{RMax}([\mathbf{z}'_m, \mathbf{z}'_l, \mathbf{z}'_t, \mathbf{z}'_{d}]),\\
 		&\mathbf{h}'_{ij} =\psi(\mathbf{W}\mathcal{S}_{ij} + \mathbf{U}(\mathbf{r}\odot[\mathbf{h}_{i,j-1}^T, \mathbf{h}_{i-1,j}^T, \mathbf{h}_{i-1,j-1}^T]^T) + \mathbf{b}), \\
 		&\mathbf{h}_{ij}     =\mathbf{z}_{l}\odot\mathbf{h}_{i,j-1}+\mathbf{z}_{t}\odot\mathbf{h}_{i-1,j}+
            		  \mathbf{z}_{d}\odot\mathbf{h}_{i-1,j-1}+\mathbf{z}_{m}\odot\mathbf{h}'_{ij},
	\end{aligned}
\end{equation}
where $\mathbf{h}_{ij}$ stands for the hidden representation at the $(i,j)$-th position, $\mathbf{z}_{m}, \mathbf{z}_{l},\mathbf{z}_{t},\mathbf{z}_{d}$ are the four gates, $\mathbf{U}$, $\mathbf{W}$, and $\mathbf{b}$ are parameters, $\sigma$ and $\psi$ stand for sigmoid and tanh function respectively, and $\textrm{RMax}$ is a function to conduct softmax on each dimension across gates,
\begin{equation}
[\mathbf{z}_\phi]_j=\frac{e^{[\mathbf{z'}_\phi]_j}}{e^{[\mathbf{z'}_m]_j}+ e^{[\mathbf{z'}_l]_j}+ e^{[\mathbf{z'}_t]_j}+ e^{[\mathbf{z'}_{d}]_j}}, \,\, \phi=m,l,t,d.
\end{equation}

The last hidden representation of 2D-GRU will be treated as the output $\mathbf{h}$, which is the bottom right one at the matrix/tensor. If you use a bi-directional 2D-GRU, both the top left one $\overrightarrow{\mathbf{h}}$ and bottom right one $\overleftarrow{\mathbf{h}}$ can be concatenated together to form the output vector, i.e.~$\mathbf{h}=[\overrightarrow{\mathbf{h}}^T,\overleftarrow{\mathbf{h}}^T]^T$.

Please note that both CNN and 2D-GRU well capture the proximity heuristics in IR. Proximity heuristic rewards a document where the matched query terms occur close to each other, which is an important factor for a good retrieval model. For CNN, if the matched query terms occur close to each other, appropriate kernels can be utilized to extract such significant matching patterns and influence the relevance score. In this way, CNN well captures the proximity heuristics. 2D-GRU can also model proximity. When there is a document where the matched query terms occur close to each other, the representation $\mathbf{h}$ will be strengthened by appropriately setting gates and other parameters. As a result, the relevance score of the document will be increased.
\subsection{Aggregation Network}
After the measurement step, we obtain a vector $\bf{h}$ to represent the local relevance between query and each query-centric context. Therefore, we need a further aggregation step to output a global relevance score. In this process, two IR principles are going to be considered in our deep architecture. One is {\em query term importance}: query terms are critical to express user\rq{}s information need and some terms are more important than others \cite{fang2004formal}. The other one is {\em diverse matching requirement}: the distribution of matching patterns can be quite different in a relevant document. 
For example, the {\em Verbosity Hypothesis} assumes that the relevance matching might be global. On the contrary, the {\em Scope Hypothesis} assumes that the relevance matching could happen in any part of a relevant document, and we do not require the document as a whole to be relevant to a query. In order to capture the two IR principles, we first conduct a query term level aggregation, in which the diverse matching requirement is taken into account. Then a term gating network is applied to capture the importance of different terms when producing the global relevance score.
\subsubsection{Query Term Level Aggregation}
In order to capture the principle of diverse relevance requirement, we need to consider the position of the corresponding query-centric context when conducting query term level aggregation. Therefore, we append each vector $\mathbf{h}$ with the position indicator to encode the position information of the corresponding query-centric context. Specifically, different position functions $g(p)$ are utilized in our aggregation network:	
\begin{flalign}
	\label{Eq.pFun}
	\text{Constant Function: } & g(p) = C , &C \in \mathbb{R}, \nonumber\\
	\text{Linear Function:  } & g(p) = (L - p) / L , &L \in \mathbb{R} ,\nonumber\\
	\text{Reciprocal Function:  } & g(p) = a / (p + b) , &a,b \in \mathbb{R}, \\
	\text{Exponential Function:  } & g(p) = a \cdot \exp(-p/b) , &a,b \in \mathbb{R} ,\nonumber
\end{flalign}
where $p$ stands for the position of the query-centric context, determined by the central word $v_p$. After this appending operation, the representation of local relevance for a query-centric context centered at word $v_p$ (denoted as $\mathbf{h}(p)$) becomes $[{\bf h}(p)^T,g(p)]^T$.

To conduct query term level aggregation, we first group $\bf{h}(p)$ with the same central word together, which stands for all the local relevances with respect to a same query term. Then RNN is used to integrate such local relevances by considering position information into consideration. That is to say, we can obtain the global relevance representation $\mathcal{T}(w_u)$ for each query term $w_u$ as follows.
\begin{equation}
	\mathcal{T}(w_u) = \overset{p_n}{\underset{p = p_1}{\mathrm{RNN}}} \bigg[\mathbf{h}^T(p), g(p) \bigg], \,\,\,p_1, \dots p_n \in \mathbb{P}(w_u),
\end{equation}
where $\mathbb{P}(w_u)$ denotes the position set of all the query-centric contexts centered on query term $w_u$.
For example, you can use GRU~\cite{GRU_cho2014learning} to capture the sequential information, which is a typical variant of RNN. 
In the experimental analysis, we show the comparisons among different position functions.

\subsubsection{Term Gating Network for Global Aggregation}
Based on query term level global relevance representations $\mathcal{T}(w_u)$, we use a term gating network (similar to that used in DRMM) to obtain the final global relevance score, by considering importances of different query terms. Specifically, we define a weight parameter $E_{w_u}$ for each query term, and linear combine all the query term level relevances as follows.
\begin{equation}
	\mathcal{F}(\mathbf{q}, \mathbf{d}) = \sum\nolimits_{w_u \in \mathbf{q}} (E_{w_u}\mathbb{I})^T \cdot \mathcal{T}(w_u),
\end{equation}
where $\mathbb{I}$ is an vector with each element set to be 1, and the dimension is set to be the same as $\mathcal{T}(w_u)$.

\subsection{Model Training}
DeepRank is an end-to-end deep neural network, which can be trained using stochastic gradient decent (SGD) methods, such as Adam~\cite{kingma2014adam}. L2 regularization and early stopping strategy~\cite{giles2001overfitting} are also used in our implementation to deal with overfitting. More implementation details will be given in the experiments.

In our experiments, we use the following pairwise hinge loss for training, since we are considering a ranking problem. For future work, we are also willing to try other pairwise losses and listwise loss functions to conduct the training process.
\begin{equation}
	\mathcal{L}(\bf{q}, \bf{d}^+, \bf{d}^-) = \max(0, 1 - \mathcal{F}(\bf{q}, \bf{d}^+) + \mathcal{F}(\bf{q}, \bf{d}^-)),
\end{equation}
where $\mathcal{L}(\bf{q}, \bf{d}^+, \bf{d}^-)$ denotes the pairwise loss between a pair of positive and negative samples $\mathbf{d}^+$ and $\mathbf{d}^-$, and $\mathcal{F}(\bf{q}, \bf{d})$ denotes the relevance score produced by DeepRank.

\section{Experiments}\label{sec:experiments}
In this section, we conduct extensive experiments to evaluate DeepRank against state-of-the-art models, including learning to rank methods, and existing deep learning methods. The experimental results on both LETOR4.0 benchmark~\cite{qin2010letor} and a large scale clickthrough data show that our model can significantly outperform all the baselines, especially when other existing deep learning methods perform much worse than learning to rank methods. Furthermore, we give detailed experimental analysis to show more insights on our model.

\subsection{Experimental Settings}
We first introduce our experimental settings, including datasets, baseline methods/implementations, and evaluation measures.
\subsubsection{Data Sets}
Since most deep models need to learn many parameters, it is not appropriate to evaluate them with small traditional retrieval dataset, such as Robust04 and ClueWeb-09-Cat-B used in \cite{DRMM}, which have only less than 300 queries.  In our experiments, we use two datasets for evaluation, i.e.~LETOR4.0~\cite{qin2010letor} and a large scale clickthrough data. The LETOR4.0 data is mainly used for comparing our model with other deep models, and the state-of-the-art learning to rank methods. The clickthrough data is larger, and we use it to compare different deep models.

LETOR4.0 dataset contains two separate data sampled from the .GOV2 corpus using the TREC 2007 and TREC 2008 Million Query track queries, denoted as MQ2007 and MQ2008, respectively. MQ2007 is a bit larger, which contains 1692 queries and 65,323 documents. While MQ2008 only contains 784 queries and 14,384 documents. Since the query number in MQ2008 is too small, which will cause the serious insufficient training problem for deep learning models, we propose to merge the training set of MQ2007 to that of MQ2008. The validation and testing set are kept unchanged. Then we form a new large data set, still denoted as MQ2008. In total, MQ2007 and MQ2008 contains 69,623 and 84,834 query-document pairs, respectively. All the baselines are conducted fairly on this new dataset for comparison. In original LETOR4.0, each query and document pair is represented as a vector containing 46 different features, which is easy for implementations of learning to rank methods. While most deep IR models (except for SQA~\cite{SQA}) do not use any handcrafted features, the raw text of query and document are used for implementation.

The large scale clickthrough data, namely ChineseClick, is collected from a commercial Chinese search engine. In the data collection process, the user is given the top 10 results for each proposed query. Clicked documents are viewed to be relevant, and the other ones are viewed as irrelevant. Since this is a Chinese dataset, we first conduct word segmentation for queries and documents. Then we apply some typical data preprocessing techniques, such as navigational queries filtering, stopping words and low frequency words (less than 50) removing. After these preprocessing, the final dataset contains 12,520 queries, 115,562 documents, and 118,835 query-document pairs. It is further divided into training/validation/testing set according to the proportion 3:1:1 of query numbers.

\subsubsection{Baseline Methods}
We adopt two types of baseline methods for comparison, including learning to rank methods and deep learning methods. 

For learning to rank approach, we compare both pairwise and listwise ranking methods. The pairwise baselines include {\em RankSVM}~\cite{RankSVM} and {\em RankBoost}~\cite{RankBoost}, which apply SVM and boosting techniques to the pairwise ranking problem, respectively. The listwise baselines include {\em AdaRank}~\cite{AdaRank} and {\em LambdaMart}~\cite{burges2010ranknet}, where {\em AdaRank} proposes to directly optimizing IR evaluation measures by boosting to obtain a ranking list, and {\em LamdaMart} uses gradient boosting for optimizing a listwise ranking loss, which is the winner of Yahoo\! Learning to Rank Challenge~\cite{chapelle2011yahoo}. Though the public results on LETOR4.0 have included RankSVM, RankBoost and AdaRank as the baselines, we are not able to conduct significant testing since the ranking list and relevance scores are missing. Therefore, we implement them on our own. Most of our results are comparable with those on LETOR4.0\footnote{\url{http://research.microsoft.com/en-us/um/beijing/projects/letor/letor4baseline.aspx}}. For RankSVM, we directly use the implementation in $\mathrm{SVM^{rank}}$~\cite{joachims2006training}. RankBoost, AdaRank and LambdaMart are implemented using RankLib\footnote{\url{https://sourceforge.net/p/lemur/wiki/RankLib/}}, which is a widely used tool in the area of learning to rank. For LETOR4.0 dataset, we use 46 dimensional standard features provided for public to evaluate the performance of learning to rank approaches. BM25-Title which calculate BM25 score between query and document title, is one of the powerful feature among these features.

For deep learning approach, we compare three existing deep IR models, i.e.~{\em DSSM} \cite{DSSM}, {\em CDSSM} \cite{CDSSM}, and {\em DRMM} \cite{DRMM}. We also compare some popular deep methods for text matching, including one representation focused method, i.e.~{\em ARC-I}~\cite{ARC-II}, and three interaction focused methods, i.e.~{\em ARC-II}~\cite{ARC-II}, {\em MatchPyramid}~\cite{MatchPyramid}, and {\em Match-SRNN}~\cite{Match-SRNN}. Implementation details are listed as follows. Firstly, all the word embeddings in these methods are learned with the Continuous Bag-of-Words (CBOW) model~\cite{mikolov2013distributed} from Wikipedia corpus, and the dimension is set to 50. In general, most deep learning baselines are applied following the original implementations. We only reduce the parameter numbers due to the relative small size of our data. For example, we use a three-layer DNN as that in the original paper of DSSM, and the node number of each layer is reduced to 100, 100, and 50. For CDSSM, we use a one-layer CNN with 50 $(1\times3)$ kernels. Therefore, a 50-dimensional vector is obtained after global pooling strategy. DRMM is directly implemented using the best configuration and the code released by~\cite{DRMM}. For ARC-I, we use a two-layer CNN with each one containing 16 $(1\times 3)$ kernels. The size of pooling in the first layer is set to 2, while the size of pooling in the last layer is set to be 2 for query representation, and 20 for document representation, respectively. For ARC-II, we use a two-layer CNN, where there are 8 kernels in each layer. The size of kernels and pooling in both layers are set to $(1\times3)$/$(3\times3)$ and $(2\times 2)$/$(2\times20)$, respectively. For MatchPyramid, we use cosine similarity to construct the word-level interaction matrix. Then a one-layer CNN is applied on the matrix with 8 $(3\times 3)$ kernels, and the size of dynamic pooling is set to $(3\times 10)$. For Match-SRNN, 2D-GRU is directly applied on the same interaction matrix, and the dimension of hidden node is set to 2. 

{\em SQA}~\cite{SQA} model combines handcraft features in the learning process, therefore it is not appropriate to directly compare it with DeepRank, which only uses automatically learned feature from raw text for ranking. For fair comparison, we delete the handcrafted features in SQA and obtain a pure deep SQA model, denoted as SQA-noFeat. Furthermore, we incorporate the handcrafted features (46 default features in LETOR4.0) into the last layer of DeepRank to obtain DeepRank-Feat, which is used to compare with SQA. For both SQA-noFeat and SQA, one-layer CNN with 50 $(1\times 3)$ kernels is used in the deep architecture. 

The DeepRank\footnote{The source code: \url{https://github.com/pl8787/textnet-release}.} for performance comparison is implemented using the following settings: the window size of query-centric context is set to 15; cosine similarity is adopted to construct the input tensor; both CNN and 2D-GRU are used in the measurement step, therefore we have two versions of DeepRank, denoted as DeepRank-CNN and DeepRank-2DGRU; the reciprocal function is used as the positional function, and GRU is adopted in the aggregation step. We also compare different settings of DeepRank for detailed analysis.

\subsubsection{Evaluation Measures}
For the evaluation on LETOR4.0, we follow the data partitions on this dataset (5-fold) and the average results are reported. While for the evaluation on ChineseClick, we train the model on the training set, tune hyper-parameters on the validation set and report the results on the testing set for comparison. Three evaluation measures~\cite{robertson2000evaluation} are used in this paper, i.e.~Precision, NDCG, and MAP. 
Furthermore, we conduct a pairwise t-test~\cite{TTest_smucker2007comparison} for significance testing with p-value lower than 0.05 (i.e.~$p$-value$\leq$ 0.05).
\subsection{Performance Comparison}
\begin{table*}
    \centering
    \caption{Performance comparison of different models on MQ2007, MQ2008 and ChineseClick. Significant performance degradation with respect to DeepRank-CNN is denoted as (-) with $\mathbf{p}$-value $\leq$ 0.05.}
	\label{Table.Experiments}

	\begin{tabular}{l l l l l l l l l l}
		\multicolumn{10}{c}{MQ2007} \\
		\hline
		Model & NDCG@1 & NDCG@3 & NDCG@5 & NDCG@10 & P@1 & P@3 & P@5 & P@10 & MAP\\
		\hline
		\hline
		\textsc{BM25-Title} & 0.358${}^-$ & 0.372${}^-$ & 0.384${}^-$ & 0.414${}^-$ & 0.427${}^-$ & 0.404${}^-$ & 0.388${}^-$ & 0.366${}^-$ & 0.450${}^-$ \\
		\hline
		\textsc{RankSVM} & 0.408${}^-$ & 0.405${}^-$ & 0.414${}^-$ & 0.442${}^-$ & 0.472${}^-$ & 0.432${}^-$ & 0.413${}^-$ & 0.381${}^-$ & 0.464${}^-$ \\
		\textsc{RankBoost} & 0.401${}^-$ & 0.404${}^-$ & 0.410${}^-$ & 0.436${}^-$ & 0.462${}^-$ & 0.428${}^-$ & 0.405${}^-$ & 0.374${}^-$ & 0.457${}^-$\\
		\textsc{AdaRank} & 0.400${}^-$ & 0.410${}^-$ & 0.415${}^-$ & 0.439${}^-$ & 0.461${}^-$ & 0.431${}^-$ & 0.408${}^-$ & 0.373${}^-$ & 0.460${}^-$ \\
		\textsc{LambdaMart} & 0.412${}^-$ & 0.418${}^-$ & 0.421${}^-$ & 0.446${}^-$ & 0.481${}^-$ & 0.444${}^-$ & 0.418${}^-$ & 0.384${}^-$ & 0.468${}^-$ \\
		\hline
		\textsc{DSSM} & 0.290${}^-$ & 0.319${}^-$ & 0.335${}^-$ & 0.371${}^-$ & 0.345${}^-$ & 0.359${}^-$ & 0.359${}^-$ & 0.352${}^-$ & 0.409${}^-$ \\ 
		\textsc{CDSSM} & 0.288${}^-$ & 0.288${}^-$ & 0.297${}^-$ & 0.325${}^-$ & 0.333${}^-$ & 0.309${}^-$ & 0.301${}^-$ & 0.291${}^-$ & 0.364${}^-$\\
		\textsc{Arc-I} & 0.310${}^-$ & 0.334${}^-$ & 0.348${}^-$ & 0.386${}^-$ & 0.376${}^-$ & 0.377${}^-$ & 0.370${}^-$ & 0.364${}^-$ & 0.417${}^-$ \\ 
		\textsc{SQA-noFeat} & 0.309${}^-$ & 0.333${}^-$ & 0.348${}^-$ & 0.386${}^-$ & 0.375${}^-$ & 0.373${}^-$ & 0.372${}^-$ & 0.364${}^-$ & 0.419${}^-$ \\ 
		\hline
		\textsc{DRMM} & 0.380${}^-$ & 0.396${}^-$ & 0.408${}^-$ & 0.440${}^-$ & 0.450${}^-$ & 0.430${}^-$ & 0.417${}^-$ & 0.388${}^-$ & 0.467${}^-$\\
		\textsc{Arc-II} & 0.317${}^-$ & 0.338${}^-$ & 0.354${}^-$ & 0.390${}^-$ & 0.379${}^-$ & 0.378${}^-$ & 0.377${}^-$ & 0.366${}^-$ & 0.421${}^-$\\ 
		\textsc{MatchPyramid} & 0.362${}^-$ & 0.364${}^-$ & 0.379${}^-$ & 0.409${}^-$ & 0.428${}^-$ & 0.404${}^-$ & 0.397${}^-$ & 0.371${}^-$ & 0.434${}^-$ \\ 
		\textsc{Match-SRNN} & 0.392${}^-$ & 0.402${}^-$ & 0.409${}^-$ & 0.435${}^-$ & 0.460${}^-$ & 0.436${}^-$ & 0.413${}^-$ & 0.384${}^-$ & 0.456${}^-$  \\ 
		\hline
		\textsc{DeepRank-2DGRU} & 0.439 & 0.439 & 0.447 & 0.473 & \textbf{0.513} & 0.467 & 0.443 & 0.405 & 0.489 \\ 
		\textsc{DeepRank-CNN} & \textbf{0.441} & \textbf{0.447} & \textbf{0.457} & \textbf{0.482} & 0.508 & \textbf{0.474} & \textbf{0.452} & \textbf{0.412} & \textbf{0.497} \\ 
		\hline
\hline
		\textsc{SQA} & 0.423 & 0.432 & 0.442 & 0.466 & 0.491 & 0.463 & 0.443 & 0.404 & 0.479 \\ 
		\textsc{DeepRank-CNN-Feat} & 0.446 & 0.457 & 0.462 & 0.489 & 0.518 & 0.483 & 0.451 & 0.412 & 0.502 \\ 
		\hline
		\\
		
		\multicolumn{10}{c}{MQ2008} \\
		\hline
		Model & NDCG@1 & NDCG@3 & NDCG@5 & NDCG@10 & P@1 & P@3 & P@5 & P@10 & MAP\\
		\hline
		\hline
		\textsc{BM25-Title} & 0.344${}^-$ & 0.420${}^-$ & 0.461${}^-$ & 0.220${}^-$ & 0.408${}^-$ & 0.381${}^-$ & 0.337${}^-$ & 0.245${}^-$ & 0.465${}^-$ \\
		\hline
		\textsc{RankSVM} & 0.375${}^-$ & 0.431${}^-$ & 0.479${}^-$ & 0.229 & 0.441${}^-$ & 0.390${}^-$ & 0.348${}^-$ & 0.249 & 0.478${}^-$ \\
		\textsc{RankBoost} & 0.381 & 0.436${}^-$ & 0.477${}^-$ & 0.231 & 0.455 & 0.392${}^-$ & 0.347${}^-$ & 0.248 & 0.481${}^-$ \\
		\textsc{AdaRank} & 0.360${}^-$ & 0.422${}^-$ & 0.462${}^-$ & 0.222 & 0.430${}^-$ & 0.384${}^-$ & 0.339${}^-$ & 0.247${}^-$ & 0.468${}^-$ \\
		\textsc{LambdaMart} & 0.378 & 0.437${}^-$ & 0.477${}^-$ & 0.231 & 0.446${}^-$ & 0.398 & 0.348${}^-$ & 0.251 & 0.478${}^-$ \\
		\hline
		\textsc{DSSM} & 0.286${}^-$ & 0.336${}^-$ & 0.378${}^-$ & 0.178${}^-$ & 0.341${}^-$ & 0.307${}^-$ & 0.284${}^-$ & 0.221${}^-$ & 0.391${}^-$ \\
		\textsc{CDSSM} & 0.283${}^-$ & 0.331${}^-$ & 0.376${}^-$ & 0.175${}^-$ & 0.335${}^-$ & 0.302${}^-$ & 0.279${}^-$ & 0.222${}^-$ & 0.395${}^-$ \\
		\textsc{Arc-I} & 0.295${}^-$ & 0.363${}^-$ & 0.413${}^-$ & 0.187${}^-$ & 0.361${}^-$ & 0.336${}^-$ & 0.311${}^-$ & 0.229${}^-$ & 0.424${}^-$ \\ 
		\textsc{SQA-noFeat} & 0.291${}^-$ & 0.350${}^-$ & 0.401${}^-$ & 0.184${}^-$ & 0.366${}^-$ & 0.332${}^-$ & 0.309${}^-$ & 0.231${}^-$ & 0.416${}^-$ \\ 
		\hline
		\textsc{DRMM} & 0.368${}^-$ & 0.427${}^-$ & 0.468${}^-$ & 0.220${}^-$ & 0.437${}^-$ & 0.392${}^-$ & 0.344${}^-$ & 0.245${}^-$ & 0.473${}^-$\\
		\textsc{Arc-II} & 0.299${}^-$ & 0.340${}^-$ & 0.394${}^-$ & 0.181${}^-$ & 0.366${}^-$ & 0.326${}^-$ & 0.305${}^-$ & 0.229${}^-$ & 0.413${}^-$ \\ 
		\textsc{MatchPyramid} & 0.351${}^-$ & 0.401${}^-$ & 0.442${}^-$ & 0.211${}^-$ & 0.408${}^-$ & 0.365${}^-$ & 0.329${}^-$ & 0.239${}^-$ & 0.449${}^-$ \\ 
		\textsc{Match-SRNN} & 0.369${}^-$ & 0.426${}^-$ & 0.465${}^-$ & 0.223${}^-$ & 0.432${}^-$ & 0.383${}^-$ & 0.335${}^-$ & 0.239${}^-$ & 0.466${}^-$ \\ 
		\hline
		\textsc{DeepRank-2DGRU} & 0.391 & 0.436 & 0.480 & 0.236 & 0.462 & 0.395 & 0.354 & 0.252 & 0.489 \\ 
		\textsc{DeepRank-CNN} & \textbf{0.406} & \textbf{0.460} & \textbf{0.496} & \textbf{0.240} & \textbf{0.482} & \textbf{0.412} & \textbf{0.359} & \textbf{0.252} & \textbf{0.498} \\ 
		\hline
\hline
		\textsc{SQA} & 0.402 & 0.454 & 0.493 & 0.236 & 0.485 & 0.411 & 0.362 & 0.254 & 0.496 \\ 
		\textsc{DeepRank-CNN-Feat} & 0.418 & 0.475 & 0.507 & 0.248 & 0.497 & 0.422 & 0.366 & 0.255 & 0.508 \\ 
		\hline
		\\		
		\multicolumn{10}{c}{ChineseClick} \\
		\hline
		Model & NDCG@1 & NDCG@3 & NDCG@5 & NDCG@10 & P@1 & P@3 & P@5 & P@10 & MAP\\
		\hline
		\hline
		\textsc{BM25} & 0.200${}^-$ & 0.320${}^-$ & 0.412${}^-$ & 0.280${}^-$ & 0.200${}^-$ & 0.174${}^-$ & 0.169${}^-$ & 0.152 & 0.373${}^-$ \\
		\hline
		\textsc{Arc-I} & 0.208${}^-$ & 0.359${}^-$ & 0.451${}^-$ & 0.286${}^-$ & 0.208${}^-$ & 0.193${}^-$ & 0.180${}^-$ & 0.153 & 0.393${}^-$ \\ 
		\textsc{SQA-noFeat} & 0.232${}^-$ & 0.368${}^-$ & 0.458${}^-$ & 0.292${}^-$ & 0.232${}^-$ & 0.194${}^-$ & 0.180${}^-$ & 0.153${}^-$ & 0.403${}^-$ \\ 
		\hline
		\textsc{DRMM} & 0.218${}^-$ & 0.346${}^-$ & 0.442${}^-$ & 0.288${}^-$ & 0.218${}^-$ & 0.185${}^-$ & 0.177${}^-$ & 0.153 & 0.392${}^-$\\
		\textsc{Arc-II} & 0.190${}^-$ & 0.329${}^-$ & 0.430${}^-$ & 0.283${}^-$ & 0.190${}^-$ & 0.180${}^-$ & 0.177${}^-$ & 0.153 & 0.373${}^-$ \\ 
		\textsc{MatchPyramid} & 0.204${}^-$ & 0.342${}^-$ & 0.436${}^-$ & 0.285${}^-$ & 0.204${}^-$ & 0.184${}^-$ & 0.178${}^-$ & 0.153 & 0.384${}^-$ \\ 
		\textsc{Match-SRNN} & 0.218${}^-$ & 0.360${}^-$ & 0.456${}^-$ & 0.295${}^-$ & 0.218${}^-$ & 0.190${}^-$ & 0.181${}^-$ & 0.153 & 0.399${}^-$ \\ 
		\hline
		\textsc{DeepRank-2DGRU} & \textbf{0.244} & 0.382 & 0.473 & \textbf{0.299} & \textbf{0.244} & \textbf{0.205} & \textbf{0.185} & 0.153 & 0.415 \\ 
		\textsc{DeepRank-CNN} & 0.242 & \textbf{0.386} & \textbf{0.476} & 0.298 & 0.242 & 0.202 & \textbf{0.185} & 0.153 & \textbf{0.416} \\ 
		\hline
	\end{tabular}
\end{table*}
The performance comparison results of DeeRank against baseline models are shown in Table~\ref{Table.Experiments}. 

\subsubsection{Performance Comparison on LETOR 4.0}

From the results on MQ2007 and MQ2008, we can see that: 
1) None of existing deep learning models could perform comparably with learning to rank methods. Some of them are even worse than BM25. The results tell us that the automatically learned features in existing deep learning models are not better than traditional extracted ones, though they are using more complex models for training. Someone may argue that the experimental findings are inconsistent with previous studies that DSSM and CDSSM can significantly outperform traditional retrieval models, as stated in~\cite{DSSM} and~\cite{CDSSM}. The reason lies in that LETOR4.0 is much smaller than the clickthrough data used in~\cite{DSSM} and~\cite{CDSSM}. In the following experiments on ChineseClick, we can see that all the deep models perform better than BM25. Therefore, deep models usually need more data for optimization, which is also the reason why we do not use Robust04 and ClueWeb-09-CAt-B used in~\cite{DRMM} for evaluation. 
2) As for the comparisons between these deep models, interaction focused ones such as DRMM, ARC-II, MatchPyramid and Match-SRNN perform much better than representation focused ones such as DSSM, CDSSM, ARC-I, and SQA-noFeat. This is consistent with the understanding that interaction signals are much more important than the semantic representation of query/document in IR, as described in~\cite{DRMM}. Furthermore, DRMM performs the best among all the deep learning baseline methods. This is because DRMM further incorporate IR characteristics into their architecture, which indicate the importance of capturing IR intrinsics in the architecture design process. 
3) Our DeepRank not only significantly outperforms the deep learning baselines, but also significantly improves the results of learning to rank methods, even only use the query and document raw text data. 
For example, the improvement of DeepRank-CNN against the best deep learning baseline (i.e.~DRMM) on MQ2007 is 16.1\% w.r.t. NDCG@1, 12.9\% w.r.t. P@1, and 6.4\% w.r.t. MAP, respectively; while the improvement of DeepRank-CNN against the best learning to rank method (i.e.~LambdaMart) on MQ2007 is 7.0\% w.r.t. NDCG@1, 5.6\% w.r.t. P@1, and 6.2\% w.r.t. MAP, respectively. The results indicate that by appropriately modeling relevance, deep learning approach can significantly outperform learning to rank approach for IR application.
4) Though SQA has used both automatically learned features and handcrafted features, the performance cannot compare with DeepRank by using only automatically learned features for ranking. If we incorporate handcrafted features into DeepRank, the performance will be further improved, as shown in DeepRank-CNN-Feat. The results demonstrate the superiority of our deep architecture. 

\subsubsection{Performance Comparison on ChineseClick}

The ChineseClick data is used to compare DeepRank with other deep learning methods. We do not include the DSSM and CDSSM baselines. That is because DSSM and CDSSM are specially designed for English data, and letter-trigram is used as the input of neural network, which is not applicable for Chinese data. If we are using the whole word embedding as the input, CDSMM will become the same as ARC-I. Therefore, we omit CDSSM and directly report the results of ARC-I for comparison. The results show that deep learning baselines perform comparably with BM25, some are even better. That is because we are using a larger data, the training of deep models become more sufficient and the performances are improved. Our DeepRank still performs the best. The improvement against the BM25 is about 21.0\% w.r.t. NDCG@1, and 11.5\% w.r.t. MAP. While the improvement against the best deep learning baseline (i.e.~Match-SRNN) is about 11.0\% w.r.t. NDCG@1, and 4.3\% w.r.t. MAP.

From the above results, we conclude that DeepRank significantly improves the results of relevance ranking, with architecture specially designed to model human's relevance generation process.

\subsection{Detailed Analysis of DeepRank}\label{discussion}
\begin{table}
    \centering
    \caption{Performance comparisons of DeepRank with different settings on MQ2007.}
	\label{Table.analysis}

	\begin{tabular}{l r r r}
		\hline
		Model & NDCG@1 & NDCG@5 & MAP\\
		\hline
		\hline
		\textsc{DeepRank-W1} & 0.422 & 0.425 & 0.470 \\ 
		\textsc{DeepRank-W7} & 0.426 & 0.444 & 0.490 \\ 
		\textsc{DeepRank-W11} & 0.438 & \textbf{0.458} & 0.495 \\ 
		\textsc{DeepRank-W15} & \textbf{0.441} & 0.457 & \textbf{0.497} \\ 
		\textsc{DeepRank-W19} & 0.430 & 0.453 & 0.496 \\ 
		\textsc{DeepRank-W23} & 0.441 & 0.455 & 0.496 \\ 
		\hline
		\textsc{DeepRank-}$\mathcal{S}_I^{\text{ind}}$ & 0.416 & 0.427 & 0.473 \\ 
		\textsc{DeepRank-}$\mathcal{S}_I^{\text{cos}}$ & 0.411 & 0.430 & 0.479 \\ 
		\textsc{DeepRank-}$\mathcal{S}_R$ & 0.425 & 0.439 & 0.482 \\ 
		\textsc{DeepRank-}$\mathcal{S}_{IR}^{\text{cos}}$ & \textbf{0.441} & \textbf{0.457} & \textbf{0.497} \\ 
		\hline
		\textsc{DeepRank-DNN} & 0.383 & 0.414 & 0.471 \\ 
		\textsc{DeepRank-2DGRU} & 0.440 & 0.447 & 0.489 \\ 
		\textsc{DeepRank-CNN} & \textbf{0.441} & \textbf{0.457} & \textbf{0.497} \\ 
		\hline
		\textsc{DeepRank-Const} & 0.384 & 0.419 & 0.473 \\ 
		\textsc{DeepRank-Linear} & 0.431 & 0.445 & 0.492 \\ 
		\textsc{DeepRank-Exp} & \textbf{0.441} & 0.454 & 0.494 \\ 
		\textsc{DeepRank-Recip} & \textbf{0.441} & \textbf{0.457} & \textbf{0.497} \\ 
		\hline
	\end{tabular}
\end{table}

DeepRank is such a flexible deep architecture that different parameter settings and neural networks can be used in the detection, measurement, and aggregation steps. Some of these settings may largely influence the final ranking performances. Therefore, we conduct a detailed analysis on MQ2007 to show the comparisons of DeepRank with different settings, with expect to give some insights for implementation. Specifically, we analyze four factors, i.e.~window size of query-centric context in the detection step, input tensor in the measurement step, neural network in the measurement step, positional function in the aggregation step. We change one factor of the above DeepRank-CNN each time to conduct the comparisons.
\subsubsection{Impact of Different Window Sizes of Query-Centric Context} The window size of query-centric context determines the scope of local relevance in the human judgment process. With a small window size, users would determine local relevance with less effort since contexts are short, but it is easy to introduce ambiguity due to limited context information. When window size is large, there are sufficient contexts to facilitate the precise local relevance judgment, but the cost is also increased and many noises may influence the judgment. We conduct an experiment to compare different window sizes of query-centric context, varying in the range of 1, 7, 11, 15, 19 and 23. The results listed at the top of Table~\ref{Table.analysis} show that the performances of DeepRank first increase and then become stable, with the increase of window size. The best performance is obtained with window size up to 11/15 (w.r.t different evaluation measures). Therefore, with considering the computational complexity, we suggest to use a comparable medium window size in real application, and the exact number need to be tuned considering averaged document length, query length, and data size.
\subsubsection{Impact of Different Input Tensors}
In order to capture both word representations of query/query-centric context and their interactions, we propose to construct a three-order tensor $\mathcal{S}$ as the input of the measure network. Here, we compare four different settings of tensor. $\mathcal{S}_I^{\text{ind}}$ and $\mathcal{S}_I^{\text{cos}}$ stand for the case when we use indicator or cosine function to construct the interaction matrix, and omit the other two matrices in the tensor. $\mathcal{S}_R$ stands for the case that only word representations of query and query-centric context is considered in the tensor, i.e.~interaction matrix is ignored. $\mathcal{S}_{IR}^{\text{cos}}$ stands for the case when we use the three-order tensor, which is exactly the DeepRank we used in the performance comparisons. From the results listed in the second row of Table~\ref{Table.analysis}, we can see the performances are improved when more information is modeled in the tensor. Therefore, both word representations of query/query-centric context and word-level interactions are important to the relevance judgement.
\subsubsection{Impact of Different Measure Networks}
The measure network is adopted to determine the relevance between query and a detected query-centric context. In the model section, we demonstrate how to use CNN and 2D-GRU to conduct such measurement, mainly because these two kinds of neural networks have the ability to capture the proximity heuristics. Of course, you can also use other deep learning architectures, such as DNN. In this section, we conduct experiments on MQ2007 to compare the three different versions of DeepRank, denoted as DeepRank-DNN, DeepRank-CNN, and DeepRank-2DGRU. The experimental results in the third row of Table~\ref{Table.analysis} show that DeepRank-CNN and DeepRank-2DGRU perform much better than DeepRank-DNN. The reason lies in that CNN and 2D-GRU both have the ability to model the proximity heuristics, while DNN cannot because it is position sensitive, which is contradict with the position independent proximity heuristic.
\subsubsection{Impact of Different Position Functions}
As described in the aggregation network, different kinds of position functions can be used to model the position importance. Here we compare DeepRank with four different position functions, i.e. {\em Constant}, {\em Linear}, {\em Reciprocal}, and {\em Exponential} functions, denoted as DeepRank-Const, DeepRank-Linear, DeepRank-Recip and DeepRank-Exp, respectively. The results listed in the fourth row of Table~\ref{Table.analysis} show that DeepRank-Recip is the best, while DeepRank-Const is the worst. As for the other two functions, DeepRank-Exp perform comparable with DeepRank-Recip, and DeepRank-Linear is a little worse than DeepRank-Recip and DeepRank-Exp. The results indicate that top positions are more important, which is consistent with many previous studies for relevance ranking in IR \cite{niu2012top}. As for the reason why reciprocal and exponential function performs better than linear function, we think this is because MQ2007 is extracted from GOV data, where title and abstraction information may play a dominant role in determining the relevance. Therefore, the functions with a long tail, as that in reciprocal or exponential function, will be favored. To sum up, the position function plays an important role in DeepRank, and users should pay more attention to the choice, which need to be conducted by considering the characteristics of different applications.

\subsubsection{Relations to Previous Models}
We also would like to point out that DeepRank has a close relationship with previous models, such as BM25, MatchPyramid, and Match-SRNN.
With some simplification, DeepRank can reduce to (or approximate) these models.

BM25 is a bag-of-words retrieval model that ranks a set of documents based on the query terms appearing in each document, regardless of the inter-relationships between the query terms within a document. The most common form is given as follows.
\begin{equation}
	\label{BM25_F}
	\textrm{BM25}(\mathbf{q}, \mathbf{d}) = \sum_{w \in \mathbf{q}} \textrm{IDF}(w) \cdot \frac{f(w,\mathbf{d}) \cdot (k_1+1)}{f(w,\mathbf{d}) + k_1 \cdot (1-b+\frac{b|\mathbf{d}|}{\text{avgdl}})}
\end{equation}
where $k_1$ and $b$ are the hyper parameters, $f(w,\mathbf{d})$ represents term frequency of $w$ in document $\mathbf{d}$, $\textrm{IDF}(w)$ represents inverse document frequency of $w$, $|\mathbf{d}|$ denotes the document length and $\text{avgdl}$ denotes the averaged document length.

We are going to show that a simplified DeepRank has the ability to approximate BM25 function. Firstly, the window size of query-centric context is set to 1. Then the indicator function is used to construct the input tensor, therefore word representations of query/query-centric context are ignored. In this way, exact matching signals are naturally captured, like in BM25. We can see that the output of tensor will be a 0-1 vector, with only the elements at the matched positions will be 1. If CNN is used in the measurement step, we can omit the convolution layer and directly use the pooling strategy to output the value 1; while if 2D-GRU is used, we can also output the value 1 by appropriately setting gates and parameters. As a consequence, the output of the measure network will be 1 for each query-centric context. At the aggregation step, we set the position function as a constant $g(p) = 1/|\mathbf{d}|$, and term weight as the IDF value, i.e.~$E_{w_u}=\text{IDF}(w_u)$. Therefore, the output of this DeepRank can be viewed as the following function:
\begin{equation}
	\label{DeepRank_F}
	\begin{aligned}
	\textrm{DeepRank}(\mathbf{q}, \mathbf{d})&= \sum_{w \in \mathbf{q}} \text{IDF}(w)\cdot \underset{p \in \mathbb{P}(w)}{\mathrm{RNN}} 
	\bigg[1, 1/|\mathbf{d}|
	\bigg]\\
&=\sum_{w \in \mathbf{q}} \text{IDF}(w) \cdot \mathcal{G}(f(w,{\bf d}),|\bf{}d|),
	\end{aligned}
\end{equation}
where the second equation is obtained because RNN is an accumulative process, and the function $\mathcal{G}$ is determined by the parameters in RNN, learned from the training data. The function $\mathcal{G}$ has high capacities to approximate the functions in the formula of BM25 since there are many parameters. Therefore, a simplified version of DeepRank can well approximate the BM25 model.

In addition, DeepRank has closer relationships with MatchPyramid and Match-SRNN. If we set the window size of query-centric context to be $k=|\mathbf{d}|$ and the weights of query term $w_u$ to be $1/f(w_u,\mathbf{d})$, DeepRank reduces to MatchPyramid or Match-SRNN, by using CNN or 2D-GRU as the measure network, respectively.

\section{Conclusions and Future Work}\label{sec:conclusion}
In this paper, we propose a new deep learning architecture, namely DeepRank. Firstly, a detection strategy is designed to extract query-centric contexts. A measure network is then applied to determine the local relevance between query and each query-centric context, by using CNN or 2D-GRU. Finally, an aggregation network is used to produce the global relevance score, via RNN and a term gating network. DeepRank not only well simulates the relevance generation process in human judgement, but also captures important IR characteristics, i.e.~exact/semantic matching signals, proximity heuristics, query term importance, and diverse relevance requirement. We conduct experiments on both benchmark LETOR4.0 data and a large clickthrough data. The results show that DeepRank significantly outperform learning to rank methods and existing deep IR models, when most existing deep IR models perform much worse than learning to rank methods. To the best of our knowledge, DeepRank is the first deep IR model to outperform existing learning to rank models. We also give a detailed analysis on DeepRank to show insights on parameter settings for implementation.

For future work, we plan to investigate the differences between the automatically learned representations of DeepRank and effective features used in learning to rank, which may introduce some insights for architecture design of more powerful deep IR models.

\section{Acknowledgments}
This work was funded by the 973 Program of China under Grant No. 2014CB340401, the National Natural Science Foundation of China (NSFC) under Grants No. 61232010, 61433014, 61425016, 61472401, and 61203298, and the Youth Innovation Promotion Association CAS under Grants No. 20144310 and 2016102. The authors would like to thank Chengxiang Zhai (UIUC) and Yixing Fan (ICT, CAS) for their valuable suggestions on this work, and Weipeng Chen (Sogou Inc.) for providing helps on the data processing of ChineseClick.

%
\bibliographystyle{ACM-Reference-Format}
\bibliography{sigproc_short}  
%
%
\newpage
\appendix 
\section{Data Preprocessing}
For pre-processing, all the words in documents and queries are white-space tokenized, lower-cased, and stemmed using the Krovetz stemmer. Stopword removal is performed on query and document words using the INQUERY stop list. Words occurred less than 5 times in the collection are removed from all the document.

\section{Query Matrix \& Context Matrix}
The constructions of query matrix and context matrix are described in detail below.
\begin{align*}\label{Eq.matchmatix1}
S_{ij}^{\text{ind}}=1 \,\,\text{if} \,\,w_i=v_j, \,\,\,S_{ij}^{\text{ind}}=0 \,\,\text{otherwise},\\
S_{ij}^{\text{cos}} = \mathbf{x_i}^T \mathbf{y_j} / (\|\mathbf{x_i}\| \cdot \|\mathbf{y_j}\|),
\end{align*}
where $\mathbf{x_i}$ and $\mathbf{y_j}$ denote the word embeddings of $w_i$ and $v_j$, respectively.
To further incorporate the word representations of query/query-centric context to the input, we extend each element of $S_{ij}$ to a three-dimensional vector ${\bf\tilde S}_{ij}$.

\begin{equation*}\label{Eq.fixmatchmatrix}
	{\bf\tilde S}_{ij}=[x_i,y_j,S_{ij}]^T
	= [(\mathbf{W^{Q}}\mathbf{x_i})^T, (\mathbf{W^{D}}\mathbf{y_j})^T, S_{ij}]^T
\end{equation*}
where $W^{Q}$ and $W^{D}$ are the linear transformations that reducing the higher dimensions of word embeddings into lower ones, for example, from 50 dimensions to 2 dimensions.

Therefore, the original matrix $\bf{S}$ will become a three-order tensor, denoted as $\mathcal{S}$. In this way, the input tensor can be viewed as a combination of three matrices, i.e.~query matrix, query-centric context matrix, and word-level interaction matrix.

\section{Code}
We have released two versions of DeepRank. The original one is TextNet (\url{https://github.com/pl8787/textnet-release}), the newer one is implemented in PyTorch (\url{https://github.com/pl8787/DeepRank_PyTorch}).

\end{document}